\documentclass{camera}
\usepackage{graphicx}  

\begin{document}

%
\title{Concluding Remarks: the Scientist that Lived Three Times}

%
\author{Luciano MAIANI}

%
\organization{Dipartimento di Fisica, Universit\`a di Roma {\it La Sapienza}\\ INFN, Sezione di Roma\\Piazzale A. Moro 5, Roma, I-00185.}

\maketitle

\begin{abstract}
The highlights of the conference: The Legacy of Bruno Pontecorvo: the Scientist and the Man, held in Roma, Universit\`a {\it La Sapienza}, 11-12 September, 2013, are summarized and illustrated. 
\end{abstract}

%
\section*{Foreword}
Bruno Pontecorvo has been a scientist that lived three times, in three great sagas of the 20$^{th}$ Century:
 the Via Panisperna boys, the Cold War life in post-war Soviet Union, the fall of real socialism and the disparition of the soviet system.

These different periods correspond, approximately, to three different focuses of his interests:
nuclear physics, particle physics, neutrinos in the Universe. He left permanent signs in all these areas.

Personally, I am very glad of the invitation to give this talk, as I had several connections with Bruno and with the Pontecorvo family. 

After 1978,  Bruno visited Roma periodically, working in an office next to mine. We discussed about Standard Theory and neutrinos.  Occasionally, I drove him to his sister's home, who lived in the other side of town, near the Vatican, and had ample time to be impressed by his personality. 

Secondly, when working at Istituto Superiore di Sanit\`a, at the beginning of my career, I established a long standing friendship with Eugenio Tabet, the son of another Bruno's sister, Giuliana, and one of the organizers of this Conference. 

Finally, I met several times Bruno's brother, Gillo, while Ludovico, Gillo's son, also a pysicist, was a good friend of my son Tito (no connection between the name of my son and the name of Bruno's second son, Tito Niels).

The Conference, opened by Jack Steinberger~\cite{steinberg} gave a brilliant picture of the scientific and human personality of Bruno Pontecorvo and put into evidence the deep influence of his work on modern particle physics. 

Indeed, I could simply limit myself, at this point, to address the reader to the very beautiful and detailed presentations we have assisted to in the last days. For those that may have not attended the talks, I shall illustrate a few points, which seem particularly significant  to me, and some personal views.

\begin{figure}[!h]
        \centering
        {
       \includegraphics[scale=0.70]{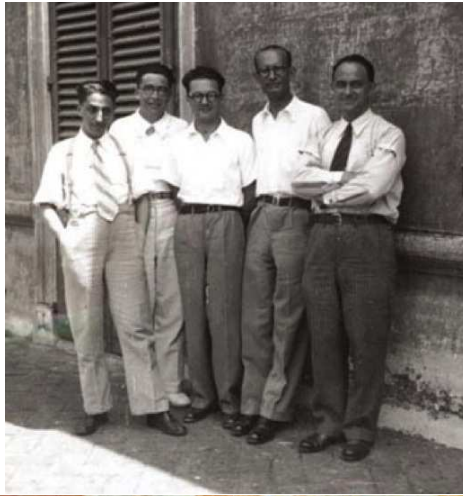}}
           { 
       \includegraphics[scale=0.70]{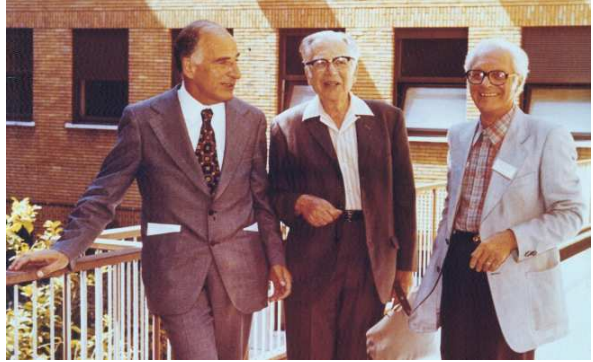}}
  	   \caption{\label{see-saw} {\footnotesize (a) Via Panisperna boys (from right): Fermi, Rasetti, Amaldi, Segr\`e, D'Agostino. The young Bruno made part of the group but does not appear, he was behind the camera taking the picture. (b) At the $70^{th}$ birthday of Edoardo Amaldi, Roma 1978. From left:  Bruno Pontecorvo, Emilio Segr\`e and Edoardo Amaldi.}}
\end{figure}

\section{The Mesotron}

Since Yukawa's theory, everybody was convinced that the quantum of nuclear forces, today called the $\pi$ meson, had to be identified with the mesotron, the particle discovered in 1937 by J. C. Street and E. C. Stevenson.

Luigi Di Lella~\cite{dilella} started his talk by recalling the experiment performed in Rome by Conversi, Pancini and Piccioni with magnetic selection of the charge sign of cosmic rays. In 1946, they found that a very large fraction of negative ÒmesotronsÓ decayed when coming to rest in Carbon, instead of being absorbed by the nucleus, as was predicted by Araki and Tomonaga.

Louis Alvarez~\cite{alvarez} commented in his Nobel lecture:
{\it ... As a personal opinion, I would suggest that modern particle physics started in the last days of World War II, when a group of young Italians, Conversi, Pancini, and Piccioni, who were hiding from the German occupying forces, initiated
a remarkable experiment}.

Fermi, Marshak and others were worried by the question: but then, where is it the pion?

Pontecorvo asked a simpler question: but then, what is it the mesotron?
and found a surprising answer: it is a second generation electron~\cite{ponte1}.

The mesotron does not arise from further subdivision of normal matter (atoms, nuclei, nucleons, atomic and nuclear forces), so I. Rabi asked:
who ordered that?   what is its role in the picture of the fundamental forces?

Pontecorvo found a provisional answer in what was called later the Universality of the Weak Interactions, embodied in the Puppi triangle of 1950~\cite{puppi} and later established by the fundamental works of Feynman and Gell-Mann~\cite{FGM1}, Gershtein and Zeldovich~\cite{GZEL}, Marshak and Sudarshan~\cite{MarSud} and, finally, Cabibbo~\cite{weakangle}.

This is the line that, eventually,  led to the electroweak unification  of Glashow, Weinberg and Salam, the GIM mechanism and to the Standard Theory of today.

Remarkably, we still do not have a plausible explanation of generations!


\section{$\mu \to e \gamma$ and Two Neutrinos}

Much later, in 1982, Pontecorvo described the antidogmatic mood he found himself in, after the Conversi, Pancini, Piccioni result, and the questions that came to his mind: 
{\it ... who says that the muon must decay in an electron and two neutrinos or into an electron and a photon? is the charge particle emitted in the muon decay an electron?...}~\cite{ponte82}. 
the same questions were, at the time, in the mind of others, in particular Steinberger, who was doing his PhD thesis with Fermi in Chicago and met Pontecorvo there~\cite{steinberg}. 

In this connection, Pontecorvo  quotes the results obtained while in Canada, in collaboration with Ted Hincks, the first one being a limit on the decay $\mu \to e \gamma$.
The story of the early limits has been reported by Di Lella~\cite{dilella}:
\begin{eqnarray}
&& R_{e\gamma}=\frac{\Gamma(\mu \to e \gamma)}{\Gamma(\mu \to all)}\nonumber \\
&&R_{e\gamma}< 0.01~({\rm Hincks ~and ~Pontecorvo,~1948,~Cosmic ~Rays});\nonumber \\
&&R_{e\gamma}< 2\cdot 10^{-5}~({\rm Lokanathan~and~Steinberger,~1955, Nevis});\nonumber \\
&&R_{e\gamma}< (1.2\pm 1.2)\cdot 10^{-6}~({\rm J.~Ashkin~et~al.,~1959,~CERN}).
\label{limit1}
\end{eqnarray}

In the Intermediate Vector Boson theory with only one type of neutrino, G.  Feinberg~\cite{Feinberg:1958zzb}
 estimated, in 1958, a value: 
\begin{equation}
R_{e\gamma}\sim 10^{-5}~({\rm W~and~one~ neutrino,~Feinberg,~1958})
\end{equation}

 To avoid inconsistency with data, several authors, around 1960, introduced two different neutrinos (among others, J. Schwinger~\cite{Schwinger:1957em}, N. Cabibbo and R. Gatto~\cite{Cabibbo:1960zz}, T. D. Lee and C. N. Yang~\cite{Lee:1960qw}). In fact, two massless neutrinos, with exact muonic and electronic number conservation, make  the amplitude to vanish exactly.

For massive neutrinos with a non-diagonal mass in the weak basis, there is a non zero amplitude proportional to $\sin\theta\cos\theta (m_1-m_2)$, a precursor of the GIM mechanism. However,  with the present limits on neutrino masses one gets an infinitesimal result.

Massive SUSY particles may change the situation, producing a detectable rate for the $e\gamma$ decay and this is why this decay is still actively searched for, together with the $\nu_\mu$ conversion process (again, see Di Lella~\cite{dilella}):
\begin{equation}
\nu_\mu +N \to e+N 
\end{equation}

A measure of the progress since the early times, and of the desert thus far encountered beyond the Standard Theory, is given by the present  upper limit produced in 2013 by the MEG Collaboration~\cite{Adam:2013mnn}:
\begin{equation}
R_{e \gamma} < 5.7 \cdot 10^{-13}
\end{equation} 

\section{Detecting Neutrinos}

The mean free path of a neutrino in normal matter is known to be of the order of light-years~\cite{bepe}.

However, Pontecorvo at Chalk River knew well that a nuclear reactor produces an absolutely extraordinary flux of neutrinos, from the neutrons which are the reactor engine. 
The probability that "some" of them could produce a visible interaction could be not so vanishingly small.

Pontecorvo proposed a radiochemical method to catch some of these neutrinos. A neutrino can trasform a nucleus of Chlorine 37 in a nucleus of Argon 37, which is beta radioactive, with half life of $34.3$ days. Argon can be chemically extracted from Chlorine, concentrated and its radioactivity measured.

The presence of Ar in Cl is a physical evidence of the existence of the neutrino and by counting the produced atoms, one could measure the interaction cross section and compare it to the prediction of the Fermi theory. Already in 1946,  Pontecorvo thinks to the neutrinos emitted by the Sun (see the talks by Giuseppe Fidecaro~\cite{fidecaro} and by Frank Close~\cite{close}).

In 1947, Pontecorvo speaks with Pauli, who encourages him. He speaks with Fermi as well, who is not excited. Perhaps Fermi thinks that it will take decades before the method can be put to work ({\it Don Quixote is not a Fermi' s hero} will comment E. Segr\`e). 
Pontecorvo starts anyway to organize his Laboratory in Canada to explore neutrino physics (1948, see~\cite{fidecaro}).

Are Neutrinos Majorana Particles?
As discussed at lenght by Fidecaro, Pontecorvo considered  two possible nuclear transitions, which in the Fermi theory are due to the trasformation of a neutron into a proton, or viceversa:
\begin{eqnarray}
&& ({\rm Cl^{37}}\to{\rm Ar^{37}}):~\nu + n \to p+e^- \\
&& ({\rm Cl^{37}}\to{\rm S^{37}}):~\bar \nu + p \to n+e^+
\end{eqnarray} 

In the Fermi theory,  neutron decay gives  an antineutrino and $\bar \nu \neq \nu$,  so the $Cl\to Ar$ method would not work with reactors, while the less favorable $Cl\to S$ would. However, at the time, the concept of lepton number was not well established  and the difference between the particle initiating the two transitions was not clear. Pontecorvo and others (including Pauli, according to Pontecorvo's recollection~\cite{ponte82}) thought that, following Majorana, it could be that $\bar \nu = \nu$.

Pontecorvo had in mind solar reactions as well, where $\nu$ is produced and the Chlorine method would work, as proved by R. Davis in the 1970s.

That the particle produced by reactors gives rise to positrons was proved by Reines and Cowan, about ten  years after Pontecorvo's proposal.
But, after maximal parity violation and $V-A$ discovery, we know that Majorana can still go through with a relatively small change (see e.g.~\cite{maia014}), the reactions being: 
\begin{eqnarray}
&& \nu(h=-1) + n \to p+e^- \; ({\rm Cl^{37}}\to{\rm Ar^{37}})\\
&& \nu (h=+1) + p \to n+e^+\;({\rm Cl^{37}}\to{\rm S^{37}}) 
\end{eqnarray}

\section{Changing Life}

In summer 1950, Pontecorvo and family  disappear from Roma, to reappear in the Soviet Union some five years later~\cite{close}.

An exceptional document was produced in the Conference, Pontecorvo's first Logbook/Notebook in Dubna, starting Nov. 1st, 1950.  Presented by Rino Castaldi~\cite{castaldi}, the Notebook shows Bruno working in particle physics and reorganizing his life as an experimental group leader in Dubna.
 There is no trace of activity in nuclear energy, no indication of a former atomic spy. 
 
\begin{figure}[!h]
        \centering 
      {
           \includegraphics[scale=0.46]{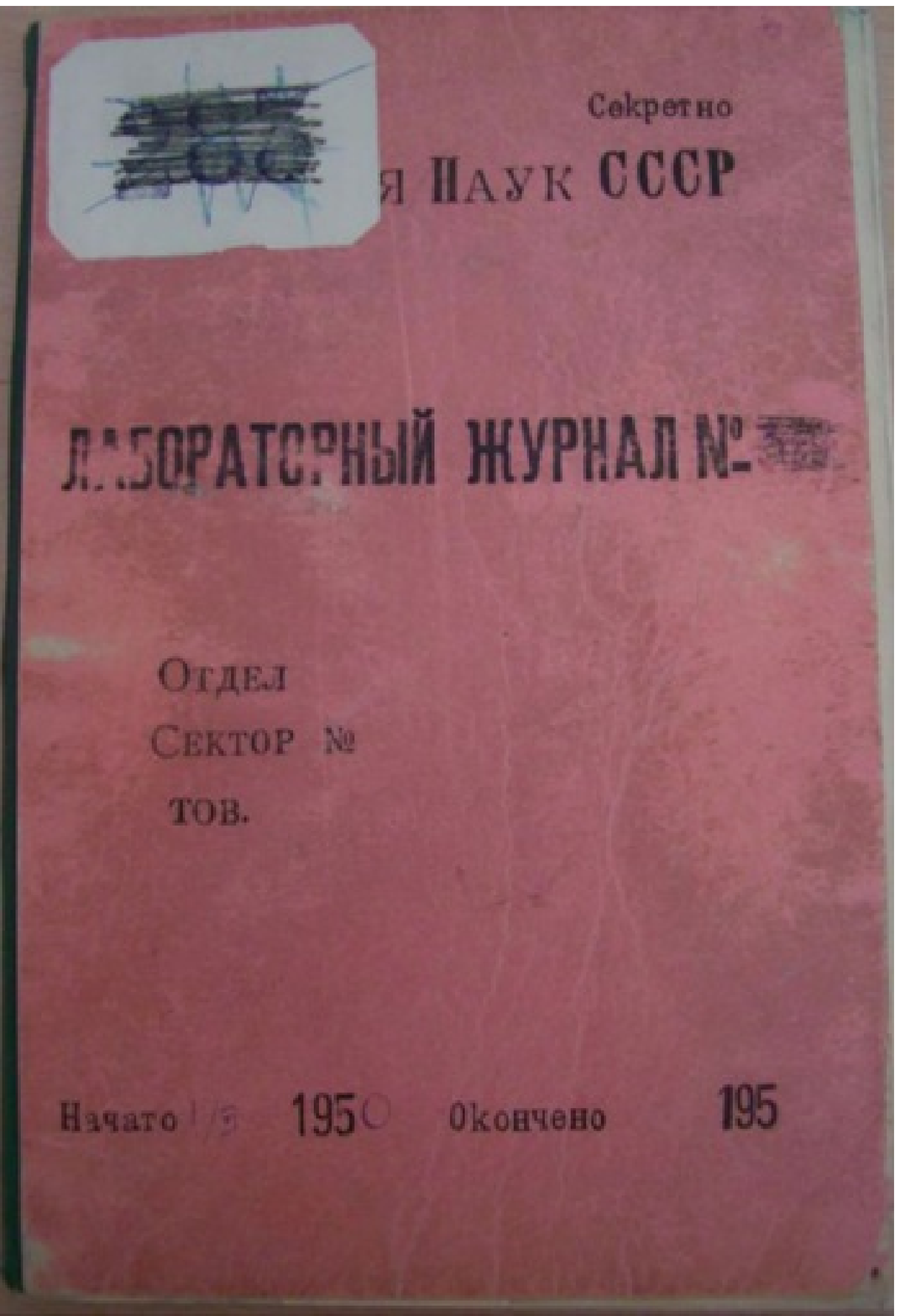}}
           { 
       \includegraphics[scale=0.30]{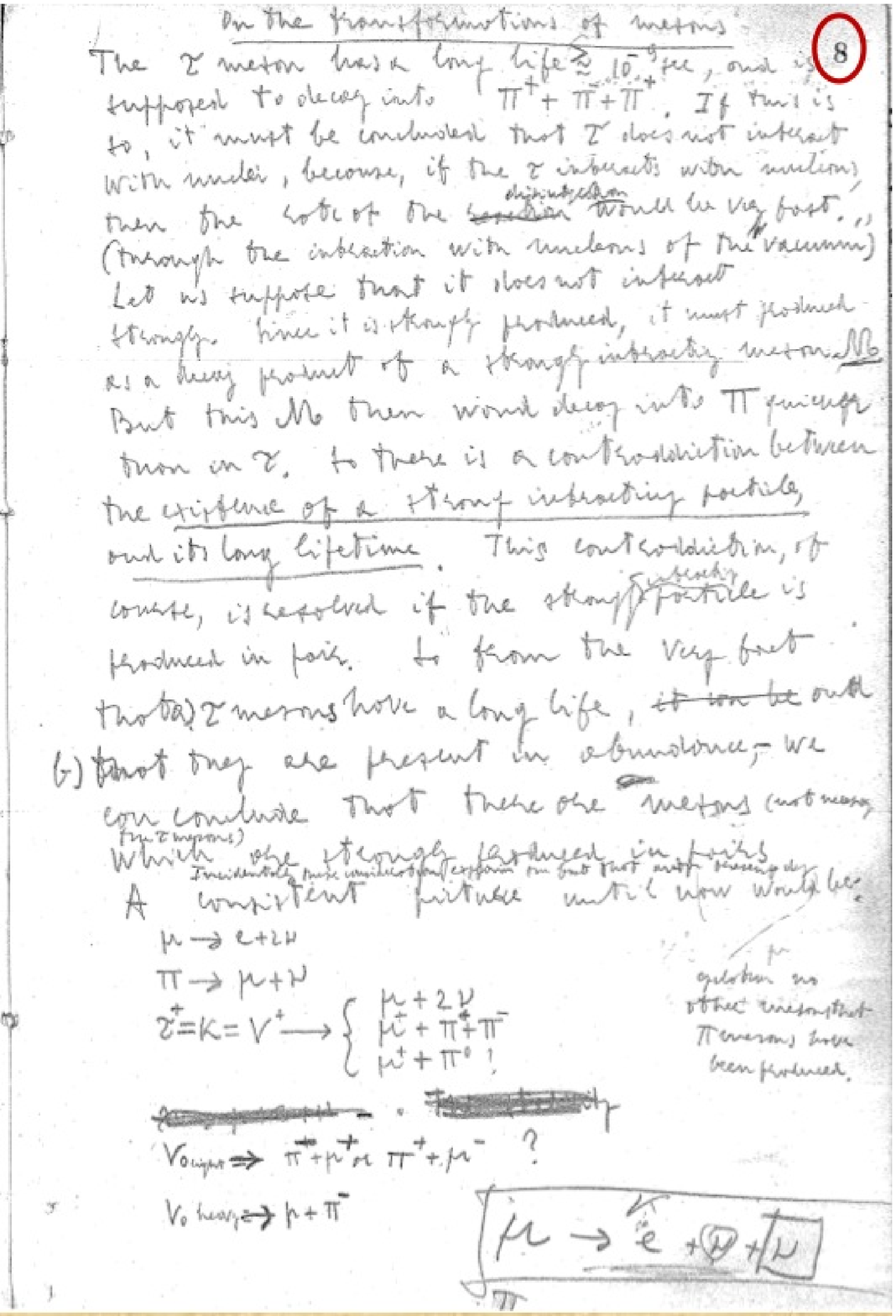}}

  	   \caption{\label{notebook} {\footnotesize The first Pontecorvo's Logbook/Notebook in Dubna, dated Nov. 1$^{st}$, 1950~\cite{castaldi}. }}
\end{figure}

 In 1951, Fermi, who had some experience in these matters, was to say about Pontecorvo's defection to the URSS  (quoted by Nadia Robotti~\cite{robotti}): {\it I do not know of course what are the reasons that prompted his alleged escape to Russia. My personal impression of his research activities has been that he did not have much interest in the atomic developments except as a tool for scientific research. In particular I do not remember any instances in which he took up with me any subject connected with atomic technology and he did not seem to have any special interest in atomic weapons
... my impression is that if he went to Russia he may not be able to contribute to their work by the things that he has learned during his connection with the Canadian and the English projects but rather through his general scientific competence}. 

Similar opinions have been expressed by Bruno's colleagues, in Italy and abroad.

The Notebook is a remarkable document, indeed. With pages after pages, written in a minute but precise writing with remarkably few cancellations (Fig.~\ref{notebook}), it reconstructs a picture of Pontecorvo building up his future activity in particle physics in the laboratory he had chosen for life. Issues in the life of the experimental physicist he was, ideas about new experiments, glimpses about his toughts on the misterious strongly produced but long-lived strange particles. 

Up to a tantalizing formula for the muon beta decay, with one neutrino encircled and the other in a box. A hint that the two neutrinos may be different?

 We leave the word to Pontecorvo himself~\cite{ponte82}. {\it I have to come back a long way (1947-1950). Several groups, among which J. Steinberger, E. Hincks and I, and others were investigating the (cosmic) muon decay. The result of the investigations was that the decaying muon emits three particles: one electron...and two neutral particles, which were called by various people in different ways: two neutrinos, neutrino and neutretto, $\nu$ and $\nu^\prime$, etc. I am saying this to make clear that for people working with muons in the old times, the question about different types of neutrinos has always been present... for people like Bernardini, Steinberger, Hincks and me Éthe two neutrino question was never forgotten}.

\begin{figure}[!h]
        \centering 
   {
    \includegraphics[scale=0.45]{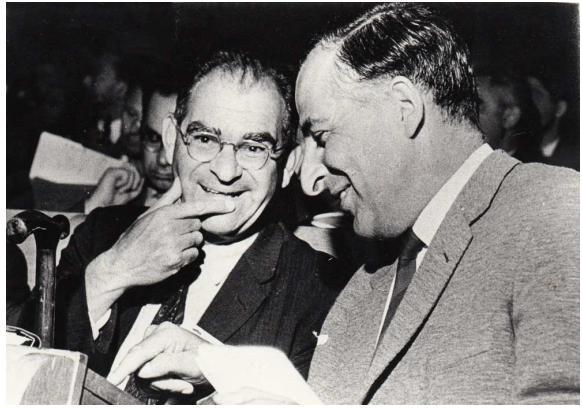}}
            \caption{\footnotesize Pontecorvo in Dubna with Vicky Weisskopf (at right).}
          \end{figure}

\section{Muon Flavor Requires its Own Neutrino} 

Towards the end of the fifties, high energy particle beams that were becoming available or were being planned, raised a new interest in neutrino physics, to study high energy neutrino interactions and to determine the identity of the neutrinos emitted with muons and/or electrons. Di Lella~\cite{dilella} has described the path that led to the discovery of two neutrinos, I may simply add a few remarks.

In Dubna, 1958, a proton beam of $800$~MeV was planned and Pontecorvo tried to figure out how to use it to see if $\nu_\mu \neq \nu_e$. He chosed the {\it beam dump} method: dumping the beam into a thick iron block, the positive pions would stop and decay, emitting essentially $\nu_\mu$s. The neutrinos have still the energy to produce either muons or electrons by scattering over a target. If $N_\mu >> N_e$, you could prove that there are two neutrinos. The principle works, indeed the experiment has been done much later in Los Alamos, to search for $\nu_\mu \to \nu_e$ oscillations~\cite{dilella}. 

Unfortunately, the proton beam in Dubna was never done. Independently, Lederman, Schwartz and Steinberger and collaborators proposed to construct a neutrino beam from the decay in flight of pions produced by the Brookhaven proton beam of $15$~GeV. The interactions of the high energy neutrinos would test wether  $\nu_\mu \neq \nu_e$, as was being proposed by Lee and Yang, among others.  

Pontecorvo's paper was known to Mel Schwartz, who acknowldged it at the end of his proposal for the neutrino beam~\cite{dilella}. 

The Brookhaven method was superior because of the high energy neutrinos and because it used the recently invented spark chambers to detect neutrino interactions. Dino Goulianos, in a recent celebration of the 1988 Nobel Prize to Lederman, Schwartz and Steinberger, quotes in verses his conversation with Mel Schwartz~\cite{dgoulia}: 
\begin{center}
\noindent {\it 
to be my student-he said-\\
remember Dino:\\
you must be really good\\
to catch a fast neutrino.\\
Oh! let me be your student\\
--- I said ---\\
don't leave me in grief\\
I promise, I'll catch the\\
first neutrino\\
with my bare teeth!
}
\end{center}
and Di Lella~\cite{dilella}: {\it ...the invention of the spark chamber (Fukui and Miyamoto, 1959) made the first high-energy neutrino experiment possible.}

\section{Neutrino Oscillations}

S. Bilenky~\cite{bilenky} gave a detailed account of Pontecorvo's trajectory towards neutrino oscillations and the latest determinations of the oscillation parameters from solar, atmpspheric and reactor neutrinos.

Prompted by the discovery of two neutrinos and by the Cabibbo theory, with the implied $d-s$ quark mixing, neutrino oscillations were accepted with great reservations, except for a few personalities. Nicola Cabibbo was one of them and he transferred to me his enthusiasm for this phenomenon. Roma was one of a few places where you could talk about neutrino oscillations without being considered a naive person. 

Besides being the main proponent, Pontecorvo had the great merit to understand that neutrino oscillations were at the basis of the so-called {\it solar neutrino deficit}, the lack of neutrinos with respect to the flux expected from the solar model, that Davis was observing in his experiment at Homestake. In the late seventies, this was an hypothesis received with even more skepticism than the idea of oscillations itself.

 Neutrino oscillations with 3 flavours, including CP and CPT violation, have been discussed by Cabibbo~\cite{cabnu} and by Bilenky and Pontecorvo~\cite{Bilenky:1978nj}.

\begin{figure}[!h]
        \centering 
      {
           \includegraphics[scale=1.0]{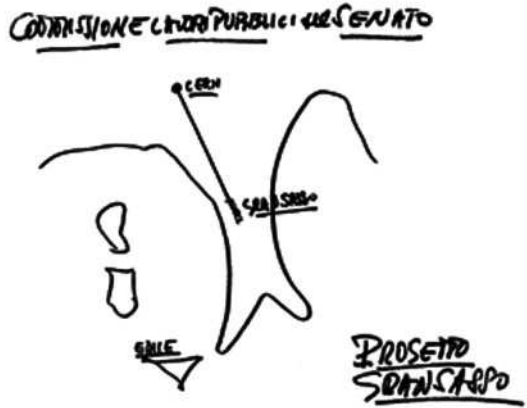}}
           { 
       \includegraphics[scale=0.40]{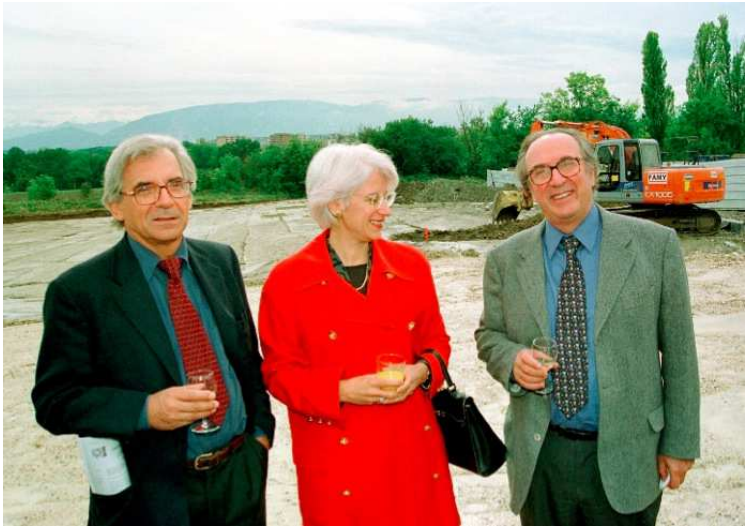}}

  	   \caption{\label{CNGS} {\footnotesize (a) The possibility of a long baseline neutrino beam from CERN to Gran Sasso was considered by A. Zichichi at the very beginning of the Gran Sasso project, as shown by this drawing shown in several occasions. As a consequence, the halls of the Gran Sasso laboratory have been oriented towards Geneva. (b) The ground breaking ceremony of the decay tunnel of pions and kaons to produce the neutrino beam was held on CERN premises in October 12, 2000. From left: A. Bettini, Vice President of INFN, Madame M. P. Bardeche, Sous-Pr\'efet de Gex, L. Maiani, CERN Director General.}}
\end{figure}

	To study neutrino oscillations, a long baseline beam from CERN to the Gran Sasso laboratory was conceived by A. Zichichi and the construction initiated under my direction of CERN, Fig.~\ref{CNGS}. Construction of CNGS (CERN Neutrinos to Gran Sasso) took place in the years 2000 to 2006. 
	
\begin{figure}[!h]
        \centering 
      {
           \includegraphics[scale=0.34]{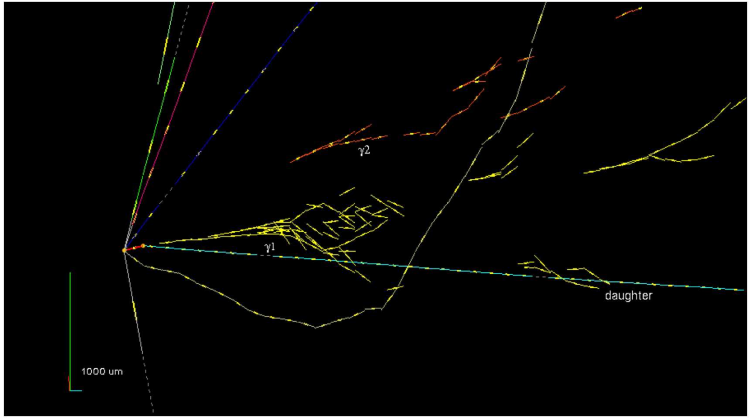}
           }         
  	   \caption{\label{tau1} {\footnotesize  The first detected Opera event. Interaction and decay vertices are joined by a red line, the decay is: $ \tau^{-} \rightarrow \rho^{-} \nu_{\tau}$~followed by $ \rho^{-}  \rightarrow \pi^{0} \pi^{-}$. Two gamma rays point to the secondary vertex, signalling the $\pi^{0}\to \gamma \gamma$ decay, the line labelled with ``daughter'' is the $\pi^{-}$~\cite{Agafonova:2010dc}.}}
\end{figure}

	The observation of atmospheric neutrinos has provided evidence for the disparition of $\mu$ neutrinos. The primary objective of CNGS was to positively observe $\tau$ production from neutrinos born at CERN as $\mu$ neutrinos and confirm the oscillation $\nu_\mu \to \nu_\tau$. The OPERA Collaboration has identified, until now, four events which show the decay in flight of a $\tau$ lepton produced in the neutrino interaction. These events are above the possible background constituted by charm production with subsequent charm semileptonic decay with a low energy muon, and provide an evidence at $4.2~\sigma$  confidence level~\cite{delellis} that neutrinos produced at CERN as $\mu$ neutrinos transform into $\tau$ neutrinos after their $730$~km journey from CERN to LNGS. Fig.~\ref{tau1} reports the first observed event~\cite{Agafonova:2010dc}.

\section{Are Neutrinos Majorana Particles (II)?} 
	
The see-saw mechanism~\cite{seesaw} makes the very small neutrino masses natural. 
If so, the flavour groups of leptons and quarks may be different.
\begin{itemize}
\item {\bf quarks}: three generations of left-handed doublets, $Q_L$, and three generations of right-handed $up$-like and $down$-like singlets,  $U_R$ and $D_R$, the flavor group is $SU(3)_Q\otimes SU(3)_{U_R} \otimes  SU(3)_{D_R}$;
\item {\bf leptons}: three generations of left-handed doublets, $\ell_L$, and three generations of right-handed charged lepton singlets, $E_R$ and of heavy, Majorana neutral lepton singlets, $N$, the flavor group is $SU(3)_\ell \otimes SU(3)_{E_R} \otimes {\cal O}(3)_N$.
\end{itemize}

According to recent speculations, started in~\cite{Froggatt:1998tj}, see also~\cite{Grinstein:2010ve}, the Yukawa couplings, which determine masses and mixing of quarks and leptons, are supposed to be the vacuum expectation values of some new scalar fields. In this scheme, their values would be determined by the minimum of some potential, which has to be invariant under the flavour group. In a recent work~\cite{Alonso:2013mca} a specific example has been worked in a model with two generations only, with the interesting result that degenerate Majorana neutrinos would imply a large neutrino mixing angle, as indeed is found to be the case. 

Criteria to find the natural minima of invariant potentials were introduced in the sixties by N. Cabibbo, in an attempt to determine theoretically the value of the weak interaction angle, and were studied with group theoretical methods by L. Michel and L. Radicati~\cite{Michel:1970mua}
and by N. Cabibbo and myself~\cite{Cabibbo:1970rza}.

With the same methods, we have been able to study the three generations case~\cite{Alonso:2013nca}, with the flavor groups given above, and confirmed the striking difference found between quarks and leptons:
\begin{itemize}
\item	{\bf quarks}: hyerarchical masses, small mixing angles;
\item {\bf charged leptons}: hyerarchical masses;
\item {\bf neutrinos}: two large angles, degenerate masses.
\end{itemize}
Note that degenerate Majorana neutrinos may correspond to non trivial mixing, as discussed in~\cite{branco}. If additional, small perturbations to the electroweak interactions are allowed, we obtain a realistic pattern of neutrino masses and mixing and provide an order of magnitude of the Majorana mass of the light neutrinos. We find:
\begin{equation}
m_\nu \sim 0.1~{\rm eV}
\end{equation}

\begin{figure}[!ht]
        \centering
       \includegraphics[scale=0.3]{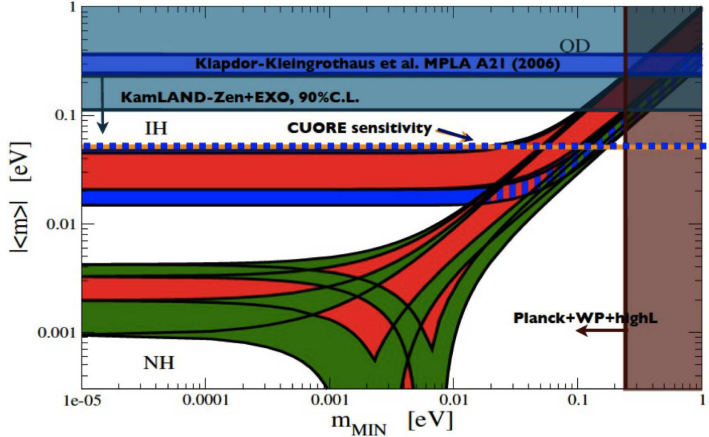}
  	   \caption{\label{PredictionsDbdecay} Neutrino masses coming from double beta decay without neutrinos for direct (lower) or inverse (upper) hierarchy. The two possibilities converge in a single line for degenerate neutrinos. Courtesy of S. Pascoli~\cite{pascoli}.  }
\end{figure}

A mass of this order would lead to a rate for neutrinoless  double beta decay not too far from the present limits, as shown in Fig.~\ref{PredictionsDbdecay}. 
Three almost degenerate neutrinos with $m_0\sim 0.1$~eV would be compatible with the recent value of the sum of neutrino  masses reported by the Planck Collaboration~\cite{Ade:2013lmv}, on the basis of cosmological data:

\begin{equation}
\sum m_\nu = 0.22 \pm 0.09~ {\rm eV}.
\end{equation}

	\section{CONCLUSIONS}
	
	The Conference has been very useful to understand the complex personality and the results obtained by Bruno Pontecorvo in his long career. 

Physics wise, Bruno has been far-sighted and has made a great school in Russia. With Franck Close, one may ask: how many Nobel Prize could have been given to Bruno? What we learned from this Conference has convinced me that at least he would have deserved one for solar neutrino oscillations.

More study is needed and warranted by the interest of the man and of the times he lived in. Along this centennial year, physicists and friends have given their memories and recollections, which are however sometime (often, some may say) wrong. 

The ball is now to professional historians. 


%

\begin{thebibliography}{99} 



\bibitem{steinberg}Steinberger  J., this Conference.
\bibitem{dilella}Di Lella  L., this Conference.
\bibitem{alvarez}Alvarez  L., Nobel Lecture, 1968.
\bibitem{ponte1}Pontecorvo  B., Phys. Rev. {\bf 72} (1947) 246.
\bibitem{puppi}Puppi  G., {\it Nuovo Cimento}, {\bf 5} (1948) 587;  Klein O., {\it Nature}, {\bf 161} (1948) 897.
\bibitem{FGM1}Feynman R.~P.~ and Gell-Mann M., {\it Phys. Rev.}, {\bf 109}  (1958) 193.
\bibitem{GZEL}Gerstein S.~S., Zeldovich Ya.~B., {\it ZhETE}, {\bf 29} (1955) 698  [JETP {\bf 2}, (1956) 576].
\bibitem{MarSud}Sudarshan E.~C.~G. and Marshak R.~E., {\it Phys. Rev.}, {\bf 109} (1958) 1860.
\bibitem{weakangle}Cabibbo  N., {\it Phys. Rev. Lett.}, {\bf 10} (1963)  531. 
\bibitem{ponte82}Pontecorvo B., {\t J. Phys. Colloques}, {\bf 43} (1982) C8-221.
\bibitem{Feinberg:1958zzb} Feinberg G.,
  {\it Phys. Rev.}, {\bf 110} (1958) 1482.
\bibitem{Schwinger:1957em}
Schwinger   J.~S.,
  {\it Ann. Phys.}, {\bf 2} (1957) 407.
\bibitem{Cabibbo:1960zz}
  Cabibbo N. and Gatto R.,
  {\it Phys. Rev. Lett.}, {\bf 5} (1960) 114.
  \bibitem{Lee:1960qw}
  Lee T.~D. and Yang C.~-N.,
  {\it Phys. Rev.}, {\bf 119} (1960) 1410.
\bibitem{Adam:2013mnn}
  Adam J. et al.  [MEG Collaboration],
  {\it Phys. Rev. Lett.}, {\bf 110} (2013) 20,  201801
  [arXiv:1303.0754 [hep-ex]].
 \bibitem{bepe}Bethe H., Peierls R., {\it Nature}, {\bf 133} (1934) 532.
  \bibitem{fidecaro}Fidecaro  G., this Conference.
    \bibitem{close}Close  F., this Conference.
  \bibitem{maia014}Maiani L., Selected Topics in Majorana Neutrino Physics, {\it Rivista del Nuovo Cimento}, to appear,  arXiv:1406.5503v1 [hep-ph].
  \bibitem{castaldi}Castaldi R., this Conference.
  \bibitem{robotti}Guerra F. and Robotti N., this Conference.
  \bibitem{dgoulia}Goulianos K., {\it Two Neutrino Experiment}, BNL 90/50/10 Celebration, June 12, 2010.
  
  
  \bibitem{bilenky}Bilenky S., this Conference.
  \bibitem{cabnu}Cabibbo N., {\it Phys. Lett.}, {\bf B72} (1978) 333.

\bibitem{Bilenky:1978nj}Bilenky S.~M. and Pontecorvo B.,
  {\it Phys. Rep.},  {\bf 41} (1978) 225.
\bibitem{Agafonova:2010dc} four $\nu_\tau$ events have been reported, until present, in:
  Agafonova N. et al.  [OPERA Collaboration],
  {\it Phys. Lett. B}, {\bf 691} (2010) 138
  [arXiv:1006.1623 [hep-ex]];  
  {\it JHEP}, {\bf 1311} (2013) 036
   [Erratum-ibid.\  {\bf 1404} (2014) 014]
  [arXiv:1308.2553 [hep-ex]]; 
  arXiv:1401.2079 [hep-ex];
  De Lellis L., seminar at LNGS, March 25, 2014. For the OPERA detector, see Dzhatdoev T.~A.~ et al.  [on behalf of the OPERA Collaboration],
  arXiv:1402.3861 [hep-ex].
\bibitem{delellis} G. De Lellis, private communication.


  \bibitem{seesaw}Gell-Mann M., Ramond P. and Slansky R.,
  Conf.\ Proc.\ C {\bf 790927} (1979) 315
  [arXiv:1306.4669 [hep-th]];  T. Yanagida, {\it Proceedings of the Workshop on Unified Theory and Baryon Number in the Universe}, KEK, Tsukuba, Japan, 1979; Glashow S. L., {\it Nato Advanced Study Institute}, {\bf B 59} (1979)  687; Mohapatra R.~N. and Senjanovic G.,
ÊÊ{\it Phys. Rev. Lett.}, {\bf 44} (1980) 912.
 
 \bibitem{Froggatt:1998tj}Froggatt C.~D. and NielsenH.~B.,
  In *Bled 1998, What comes beyond the standard model* 29-39
  [hep-ph/9905445].

\bibitem{Grinstein:2010ve}
Grinstein B., Redi M., and Villadoro G.,  {\it JHEP}, {\bf 1011} (2010) 067.

 \bibitem{Alonso:2013mca}Alonso R., Gavela M.~B., Hern\'andez D., Merlo L. and Rigolin S.,
  {\it JHEP}, {\bf 1308} (2013) 069
  [arXiv:1306.5922, arXiv:1306.5922 [hep-ph]].

\bibitem{Michel:1970mua}
  Michel L. and Radicati L.~A.,
 {\it Proc. of the Fifth  Coral Gables Conference on Symmetry principles at High Energy}, edited by B. Kursunoglu et al.,  W. H. Benjamin, Inc. New York (1965);  {\it Annals Phys.}, {\bf 66 } (1971)  758.

\bibitem{Cabibbo:1970rza}
  Cabibbo N. and Maiani L.,
 in {\it  Evolution of particle physics}, Academic Press (1970), 50, App. I.
 
 \bibitem{Alonso:2013nca}
  Alonso R., Gavela M.~B., Isidori G. and Maiani L.,
  {\it JHEP}, {\bf 1311} (2013) 187
  [arXiv:1306.5927 [hep-ph]].
  
  \bibitem{branco}Branco G. C., Rebelo M. N.,  Silva-Marcos J. I., {\it Phys. Rev. Lett.}, {\bf 82} (1999) 683; same authors with D. Wegman, arXiv:1405.5120 [hep-ph]. 
   
  \bibitem{Ade:2013lmv}
  Ade P.~A.~R.  et al.  [Planck Collaboration],
  arXiv:1303.5080 [astro-ph.CO].
  \bibitem{pascoli}~S. Pascoli, private communication.
  
  
 \end{thebibliography}
\end{document}